\UseRawInputEncoding
\documentclass[aps,twocolumn,superscriptaddress,groupedaddress]{revtex4}
\usepackage{graphicx}

\usepackage{amssymb} 
\usepackage{natbib} 
\usepackage{color}
\usepackage{amsmath}

\begin{document}
\widetext
\title{Field Induced Modulated State in the Ferromagnet PrPtAl}
\author{Christopher D. O'Neill}
\affiliation{School of Physics and CSEC, University of Edinburgh, Edinburgh, EH9 3FD, United Kingdom.}
\author{Gino Abdul-Jabbar}
\affiliation{School of Physics and CSEC, University of Edinburgh, Edinburgh, EH9 3FD, United Kingdom.}
\author{Didier Wermeille}
\affiliation{XMAS, ESRF, BP220, F-38043 Grenoble, France.}
\author{Philippe Bourges}
\affiliation{Laboratoire L\'{e}on Brillouin (UMR12 CEA-CNRS), 91191 Gif-sur-Yvette Cedex, France}
\author{Frank Kr\"uger}
\affiliation{London Centre for Nanotechnology, University College London, Gordon St., London, WC1H 0AH, United Kingdom}
\affiliation{ISIS Facility, Rutherford Appleton Laboratory, Chilton, Didcot, Oxfordshire OX11 0QX, United Kingdom}
\author{Andrew D. Huxley}
\affiliation{School of Physics and CSEC, University of Edinburgh, Edinburgh, EH9 3FD, United Kingdom.}
\date{\today}
\begin{abstract}
\noindent
\small
The theory of quantum order-by-disorder (QOBD) explains the formation of modulated magnetic states at the boundary between ferromagnetism and paramagnetism in zero field. PrPtAl has been argued to provide an archetype for this. Here, we report the phase diagram in magnetic field, applied along both the easy $a$-axis and hard $b$-axis. For field  aligned to the $b$-axis, we find that the magnetic transition temperatures are suppressed and at low temperature there is a single modulated fan state, separating an easy $a$-axis ferromagnetic state from a field polarised state.   This fan state is well explained with the QOBD theory in the presence of anisotropy and field. Experimental evidence supporting the QOBD explanation is provided by the large increase in the $T^2$ coefficient of the resistivity and direct detection of enhanced magnetic fluctuations with inelastic neutron scattering, across the field range spanned by the fan state. This shows that the QOBD mechanism can explain field induced modulated states that persist to very low temperature.  \end{abstract}
\maketitle
\indent The suppression of magnetic order by pressure $(P)$ or chemical substitution is a proven approach to discover new quantum phases of matter, such as unconventional superconductivity. In clean metallic anti-ferromagnets, the transition remains continuous as the ordering temperature is suppressed by the tuning parameter, resulting in a quantum critical point (QCP) at zero temperature. For clean metallic ferromagnets a QCP is, however, avoided in one of two ways  \cite{RevModPhys}. In the first, the transition becomes 1$^\text{st}$ order at a tri-critical point (TCP). Tuning  beyond this point, meta-magnetic transitions occur at finite field along the easy axis that give rise to wings in the  $P$-$H$-$T$ phase diagram, across which the uniform moment is discontinuous ($H$ is the magnetic field).  This mechanism arises from coupling to any bosonic mode at zero wavevector~\cite{Belitz_1997, Chubukov_2004, Rech_2006}. Examples include UGe$_2$~\cite{Pfleiderer_2002, Taufour_2010} and ZrZn$_2$~\cite{Ulharz_2004}. In the second way a modulated state is formed between the ferromagnetic (FM) and paramagnetic (PM) states~\cite{Conduit_2009, Karahasanovic_2012}. This is driven by increased  particle-hole fluctuations around the deformed Fermi surface in the modulated state, a mechanism known as quantum order-by-disorder (QOBD)~\cite{Green_2018}.\\
\indent The first reported observation of such a modulated state was in PrPtAl~\cite{Jabbar_2015}, but those measurements did not access the QCP.  A modulated state was later observed in the itinerant magnet Nb$_{1-y}$Fe$_{2+y}$ \cite{Friedemann_2018, Niklowitz_2019}, at low temperature for $y\approx 0$ between two FM states, with $y\gtrsim 0.004$ and $y\lesssim -0.012$~\cite{Friedemann_2018}.  For excess Fe ($y\gtrsim 0.004$), the modulated state is undercut by FM at low temperature, giving a behaviour  resembling that in PrPtAl.  An avoided QCP under pressure in LaCrGe$_3$~\cite{Taufour_2016, Kaluarachchi_2017}, has recently been shown to give way to short-range order, rather than long-range modulated anti-ferromagnetism~\cite{Gati_2020}.   \\
\indent For PrPtAl, neutron and resonant X-ray scattering  identified that as a function of temperature the PM to FM transition passes through two  incommensurately modulated spin density wave states, SDW1 \& SDW2.  The SDW2 state fits well with predictions for QOBD~\cite{Jabbar_2015}. However, there are some problems  describing SDW1 in terms of QOBD, that we address in the present study. For SDW1 \& SDW2,  pressure does not suppress the ordering temperature, but enhances it, so the quantum regime where the transitions occur at very low, and ultimately zero temperature, has not so far been explored. Here, we show that a field transverse to the easy axis can provide an appropriate tuning parameter to depress the transition temperatures (to zero) in PrPtAl, giving a  fan state (SDW3), that we  explain with the QOBD theory in an applied field. This vastly expands the scope over which this theory has been successfully applied to include states that extend to zero temperature, compared with SDW1 \& SDW2, which are confined to finite temperatures.  \\
\begin{figure*}[t]
\includegraphics[scale=0.35]{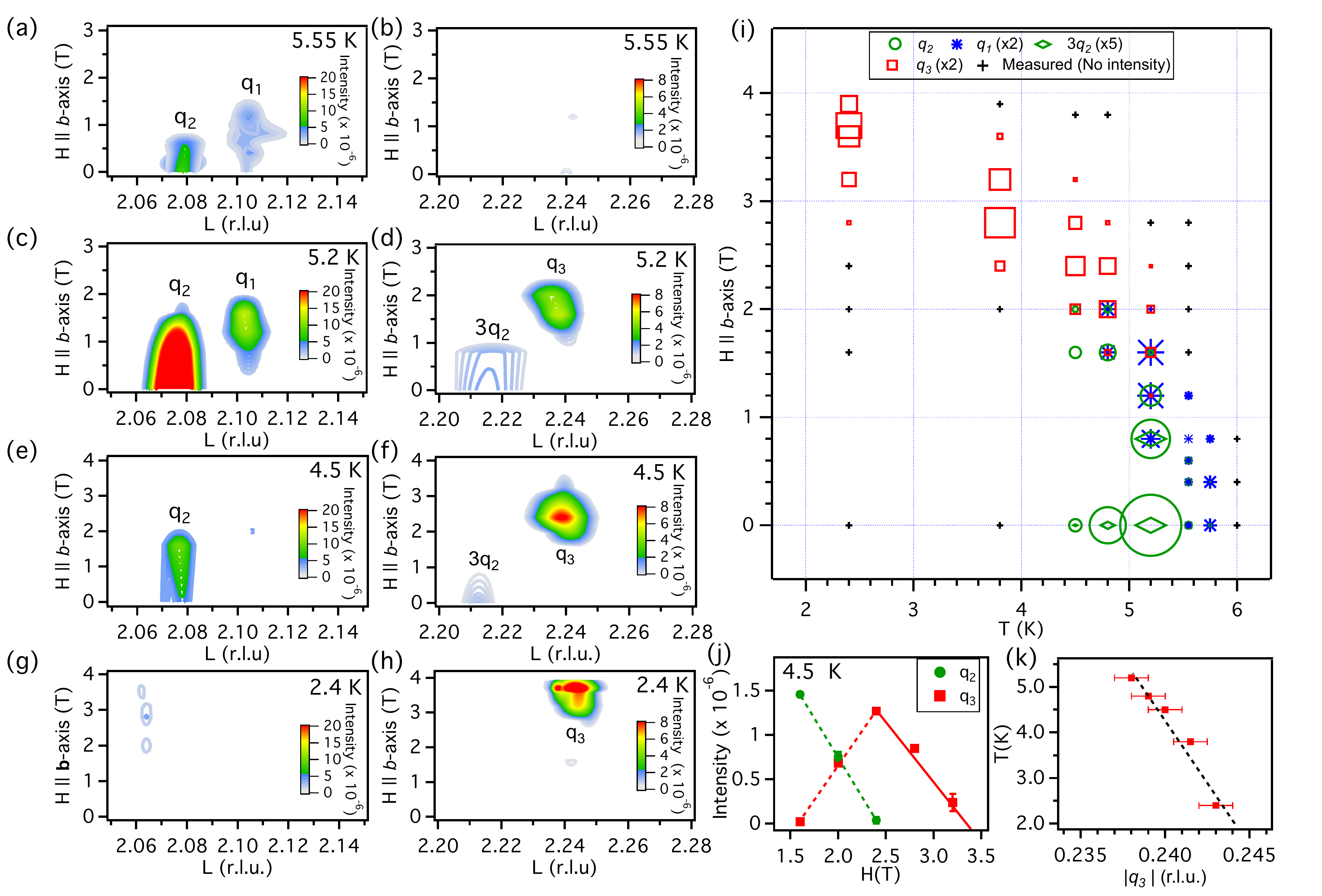}
\caption{  Colourscale images of  normalised magnetic resonant X-ray scattered-intensity as a function of field ($H$)  applied along the $b$-axis and scattering vector $(0,0,L)$ at (a) and (b) 5.55 K, (c) and (d) 5.2 K, (e) and  (f) 4.5 K and (g) and (h) 2.4 K.  Scattering at wavevectors $q_1$ and $q_2$, survives up to 2 T, before abruptly being replaced by scattering at wavevector $q_3$. (i)~The $H$-$T$ phase diagram showing the integrated intensity at $q_1$, $q_2$, $q_3$ and $3q_2$ up to 4 T.  The marker size is proportional to the integrated intensity  (scaling shown in the legend). Measured points where no intensity was found are  marked by crosses.  (j)~The variation in integrated intensity at $q_2$ and $q_3$ with $H$ at 4.5 K. (k) $T$ dependence of the magnitude of $q_3$ (the dashed line is a guide to the eye).   }
\label{Fig_1}
\end{figure*}   
\indent One of the most remarkable properties of the QOBD theory is that it explains order along magnetic hard axes~\cite{RevModPhys}. In zero field this is manifest by states SDW1 \& SDW2 with modulated moments ($m$) along both the $a$-axis (easy-axis) and $b$-axis (hard-axis) directions. The SDW3 state links the uniform states, FM and polarised PM, with different moment orientations. This provides an ideal setting for the QOBD mechanism since the difference in energy between $m\parallel\bf{a}$ and $m\parallel\bf{b}$ is low in this region.\\
\indent Mechanisms for forming SDW1 \& SDW2,   based on domain walls, a Devil's staircase generated by competing interactions, or an electronic nesting instability, have been ruled out in the previous work~\cite{Jabbar_2015}. To treat PrPtAl the QOBD model  for an itinerant system \cite{Karahasanovic_2012} was extended by the inclusion of local moments and anisotropy~\cite{Jabbar_2015}.   \\
\indent  The magnetoresistance (MR) for field along the easy $a$-axis is strongly negative in the SDW2 state with a cusp-like maximum at B=0, suggesting stronger fluctuations are present in the SDW2 state than in the FM state, supporting a QOBD based explanation for SDW2 (Supplementary Material (SM) Fig.~S2).  The MR for SDW1, however, has a peak  at low field which seems at odds with the expectation of the QOBD theory, assuming field suppresses the modulated order.  \\
\begin{figure*}
\includegraphics[scale=0.35]{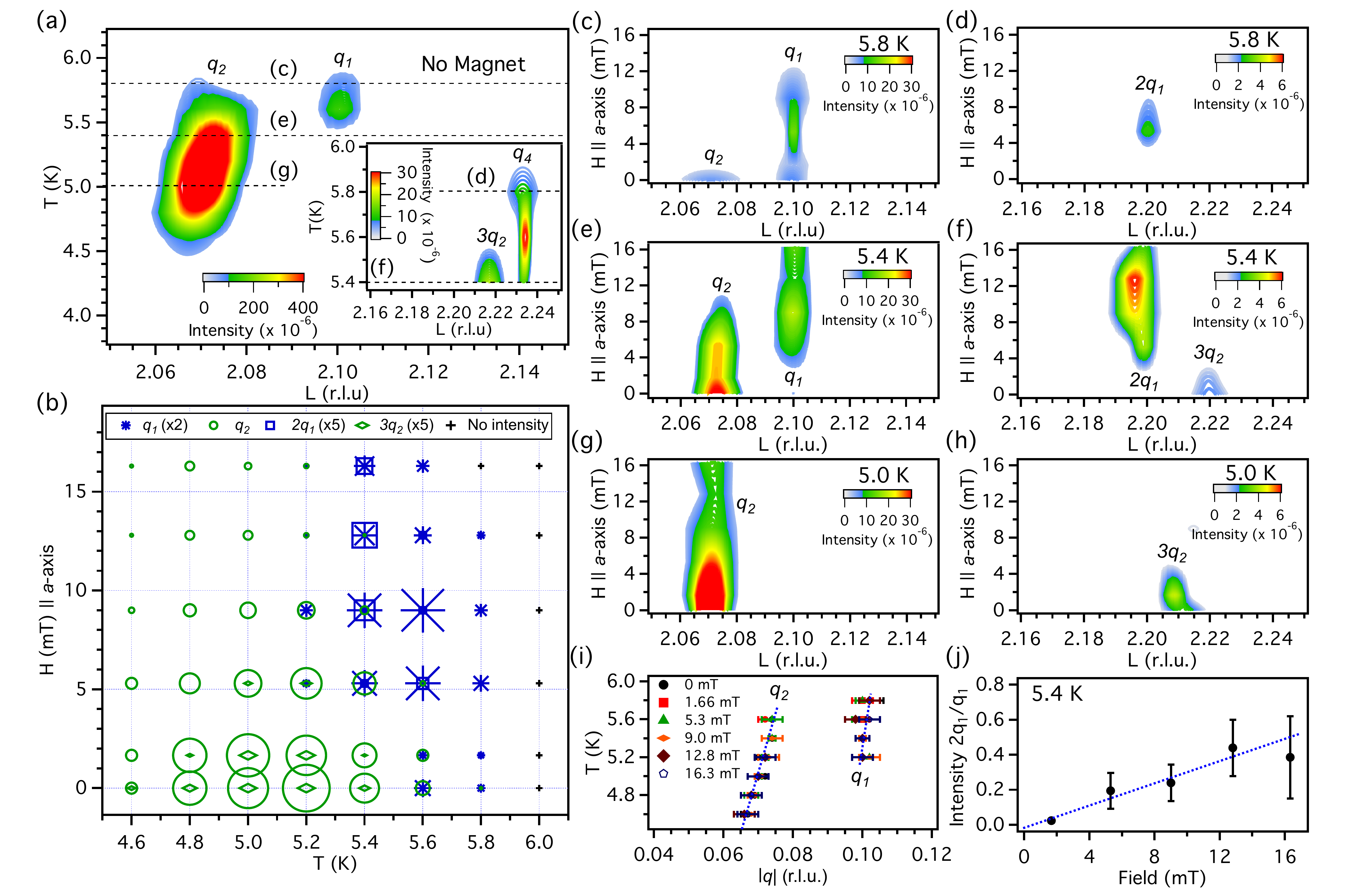}
\caption{  (a) Colourscale image of  the normalised magnetic resonant X-ray scattered intensity as a function of $T$ and reciprocal lattice co-ordinate (0,0,$L$)  in zero field (no magnet). Inset, the corresponding intensity at wavevectors $q_4$ and $3q_2$.  (b) $H$-$T$ phase diagram of the integrated intensity at $q_1$, $q_2$, $2q_1$ and $3q_2$  for $H$ applied along the $a$-axis.  The marker size is proportional to the normalised integrated intensity (scaling is shown in the legend). Measured points where no intensity is found are marked by crosses. The data at zero field in this plot are with the magnet in place.  (c)-(h) Colourscale maps of scattered intensity as a function of $H$ along $a$-axis and $(0,0,L)$ for 5.8, 5.4 and 5.0 K (the corresponding temperatures at zero field are marked by dashed lines in (a)).  (i) $T$ dependence of $q_1$ and $q_2$Ó. Points at different field superimpose. Dashed lines are guides to the eye showing that both $q_1$ and $q_2$ increase linearly with temperature. (j) The ratio of integrated intensity between the second harmonic $2q_1$ and $q_1$ at 5.4~K as a function of field.  The dashed line is a guide to the eye.  }
\label{Fig_2}
\end{figure*}
Here, we show that the amplitude of the SDW1 state is in fact initially enhanced with magnetic field before being suppressed at higher field. The increase of the MR is then perfectly consistent with QOBD. This is because in a QOBD state  the fluctuations are enhanced along with the amplitude of the order. In contrast, for non-QOBD states magnetic fluctuations are peaked at a phase transition, but are suppressed entering the ordered state.  The correlation we report  between the order and MR, therefore, provides clear evidence that both SDW1 and SDW2 are explained by the QOBD mechanism.\\
\indent  A comprehensive description of the sample and experimental methods is given in the SM.  Resonant X-ray scattering intensity  from high quality single crystals of PrPtAl, for fields along the ${b}$ and ${a}$-axis, is shown in Fig.~\ref{Fig_1} and Fig.~\ref{Fig_2}, respectively. Measurements were carried out at the 6.444~keV Pr $L_2$-edge (this energy gives access to several Brillouin zones and avoids the surface sensitivity at lower energy $M_4$ \& $M_5$ edges).  These measurements are sensitive to magnetic moments directed along $\bf{a}$.    We discuss first the low field ($<50$~mT) measurements for both axes, and then higher field measurements for $H\parallel\bf{b}$.  The overall phase diagram is shown in Fig.~\ref{Fig_4}. As found previously in zero field, SDW1 exists below $T_1=5.85\pm0.05$~K with $q_1\approx(0,0,0.10)$, accompanied by a second state with a modulation vector,  that we label here, $q_4$ $\approx(0,0,0.235)$.  Below $T_2=5.45\pm0.35$~K,  these states are replaced by SDW2 with $q_2\approx(0,0,0.07)$ and a third harmonic.     We find that the  modulation at $q_4$  is suppressed by residual fields in our magnets.    Thus, the $q_4$ intensity seen in the inset of Fig.~\ref{Fig_2}(a) (no magnet) is not present in Fig.~\ref{Fig_2}(d).   Fig.~\ref{Fig_2}(c) shows that for $H\parallel\bf{a}$, the intensity of the $q_1$ modulation of SDW1 is initially enhanced, and  is maximum at  10~mT, where it is accompanied by a second harmonic (Fig.~\ref{Fig_2}(e) and (f)). No $2q_1$ signal is induced for  field $\parallel\bf{b}$. The MR for field along the $a$-axis (SM Fig.~S2), contains a small positive maximum at $10$~mT in SDW1 that as explained above may now be understood to be a  consequence of the enhanced order in small applied field (in line with QOBD).  $\mid$$q_1$$\mid$ and $\mid$$q_2$$\mid$ increase with temperature as expected by the QOBD mechanism, shown in Fig.~\ref{Fig_2}(i). No changes of $q_1$ and $q_2$  are seen with field.\\
\indent We now discuss larger fields applied along the $b$-axis~(Fig.\ref{Fig_1}). SDW1 and SDW2 survive to $\sim 2$~T before switching to SDW3 with modulation vector $q_3\approx0.24$ (Fig.~\ref{Fig_1}(c)-(f)). The magnitude of the critical field (2~T) is comparable to the conventional anisotropy field estimated in~\cite{Jabbar_2015}.  SDW3, then extends  into the quantum regime at low temperature and high fields, which was the original target for the QOBD description.  The integrated intensities of SDW2 \& SDW3 as a function of field at 4.5 K, are shown in Fig.~\ref{Fig_1}(j).  At 1.6 T no SDW3 intensity is present,  at 2~T it co-exists with SDW2, replacing SDW2 entirely at 2.4~T consistent with  a  1$^\text{st}$ order transition. Above 2.4~T, the intensity decreases continuously with  field to zero above 3.2~T.  The linear suppression of the intensity with field suggests that the high field transition is  continuous.  This is also confirmed with  neutron scattering for $H\parallel b$-axis at 1.7~K (SM Fig. S1). The neutron scattering also confirms the uniform FM moment is suppressed when SDW3 appears  (SM Fig. S1). The resolution limited SDW peaks and suppression of FM show that the SDW3 state is distinct from FM. SDW3 could be either a polarised spiral,  a  $b$-axis fan state or an inclined plane wave state (these cannot be distinguished based on our data). A fan state,  as found in field polarised rare earth helimagnets~\cite{Herz_1978,Zverev_2014}, might be considered the most likely choice, although the mechanism driving modulated state formation is quite different.   Fig.~\ref{Fig_1}(k) shows that unlike $q_1$ and $q_2$, $q_3$ decreases with increasing temperature.  Analogous to $q_1$ and $q_2$, no significant change of $q_3$ is seen with field.   \\
\indent  MR for field along the $b$-axis is shown in Fig.~\ref{Fig_3}(a). Red markers show the transition fields seen with X-rays.  At the continuous transition to the fully polarised state, where the transverse magnetic susceptibility is expected to diverge, a local maxima exists.  The temperature range of the measurements does not permit a meaningful estimate of the power law describing the temperature dependence of the resistivity, however, the magnitude of the dependence can be estimated based on $\rho=\rho_0+AT^2$. The $A$ coefficient of resistivity as a function of field applied along the $b$-axis is shown in Fig.~\ref{Fig_3}(b),  determined from the data  between 2.2 and 4~K.  In this temperature range, SDW3 exists between $\sim 2$ and 4.2~T. The  temperature dependence of the resistivity is  seen to be enhanced in the SDW3 state.  The incommensurate modulation could provide an additional source of scattering. Following Matthiessen's rule, this would add an additional term to $\rho_0$.  The magnitude of this term would depend on the amplitude and possibly the wavevector $q_3$ of the modulation.  The SDW amplitude is suppressed with $T$, while $q_3$ changes only very modestly ($<1\%$). Thus, the overall effect of such a contribution would be to decrease  the total $T$-dependence in the fan state (the opposite of what is seen). The increased $T$ dependence observed, therefore, must reflect an increase in the DOS in SDW3 compared with FM \& PM, which confirms a key prediction of the QOBD theory.   \\
\indent Inelastic neutron scattering spectra at 1.7~K, with $H\parallel\bf{b}$, are shown in Fig.~\ref{Fig_3}(c)-(f). For PrPtAl, there are 4 Pr atoms per unit cell and the crystalline electric field (CEF) environment splits the $4f^2$ Pr$^{3+}$ ions into 9 non-magnetic singlets.  FM order is achieved by mixing singlets via an inter-site exchange interaction~\cite{Kitazawa_1998, Bleany_1963}. At 2 T, in the FM state (Fig.~\ref{Fig_3}(c)), scattering from the  excitations of the lowest excited singlet exists between 0.6-0.8~meV, consistent with previous results in zero field~\cite{Jabbar_2015}.  In  the SDW3 state,  at  3.5 and 4.5~T (Fig.~\ref{Fig_3}(d) and (e)), softening of this mode occurs at $q_3$. This softening, at nonzero $q$ and the associated energy fluctuations, require long range interactions transmitted by the itinerant electrons. Conversely the interaction with the electrons results in strong damping of the CEF levels. The broad low energy intensity at 0-0.3~meV near $q_3$, is the direct observation of this. Enhanced scattering in the same energy range is also seen close to $q=0$ (at (002)), which is a direct manifestation of an increased DOS~\cite{Lonzarich_1995}. Importantly, these strongly damped modes are also present at 3.5~T, well away from the critical field just above $\sim4.5$~T, showing that strong electronic correlations are an intrinsic attribute of the incommensurate phase and are not limited  to the critical field.    \\
\indent    We now compare our results  to the predictions of the QOBD model  with magnetic anisotropy in an applied field.  This model is described in detail in the SM. In PrPtAl,  the local CEF environments are tilted in the $a$-$c$ plane.  This means  moments in the $\bf{a}$-direction also imply an implicit AF moment component along ${\bf c}$ within the unit cell. For simplicity, we omit mention of the $c$-axis moments in the following. In previous QOBD calculations, only a helimagnet spiral state for SDW2 and uniform ferromagnetism were considered~\cite{Jabbar_2015}.  The  order parameter for the moments in the   helimagnet  is 
\begin{equation}
\mathbf{m}_\textrm{helix}(\mathbf{r}) = \left( \begin{array}{c} m_a\cos[qz + \phi(z)] \\   m_b\sin[qz + \phi(z) +\epsilon]    \end{array} \right)
\end{equation}
with $\phi(z) = -\delta_{1a} \sin(qz) + \delta_{1b} \cos(qz) - \delta_2\sin(2qz)$ where $q$  is the primary modulation vector, which is along $z$ (${c}$-axis).
Previously  only  $\delta_2 \ne 0$ was considered, which accounts for the deformation of the SDW2 state away from an evenly pitched spiral in response to the CEF and generates a third-order harmonic at $3q$ (Fig.~\ref{Fig_4} SDW2).  The term $\epsilon$ switches between a spiral for $\epsilon=0$ and an inclined plane wave for $\epsilon=\pm \pi/2$. This parameter has not been determined experimentally.

\begin{figure}[t]
\includegraphics[scale=0.55]{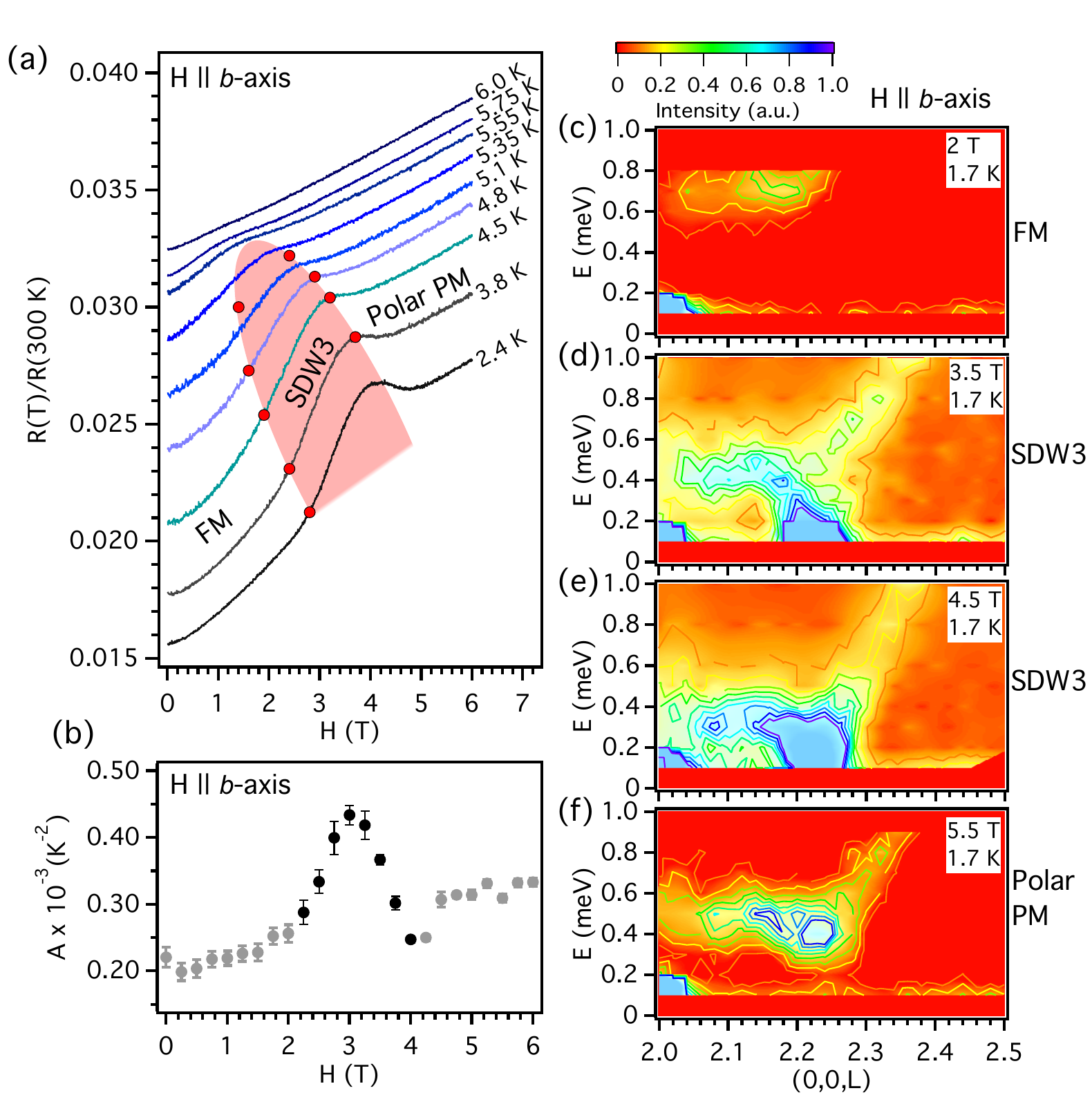}
\caption{(a) The MR, normalised to the zero field resistivity value at 300~K, for field applied along the $b$-axis.  The region where SDW3 exists is shaded red, with markers showing the transition field seen in X-ray scattering. (b) The $A$ coefficient of resistivity for Fermi liquid fits between 4-2.2 K as a function of field.  Black markers correspond to the region where  SDW3 exists within this temperature range.  (c)-(f)~Colourscale images of  inelastic neutron scattered intensity (in arbitrary units) as a function of energy transfer and reciprocal lattice co-ordinate $(0,0,L)$ at 1.7~K for different fields applied along the $b$-axis. These show the dispersion of the lowest energy magnetic excitation at  (c)~2~T in the FM state, (d) 3.5~T and (e)~4.5~T in the  SDW3 state and (f) 5.5~T in the polar PM state.   }
\label{Fig_3}
\end{figure} 

\indent The $\delta_{1a,b}$ terms tilt moments towards a field along the $a,b$-axis. These terms give a second-order harmonic.  Experimentally the SDW2 state  resists polarising in a field ($\delta_1$ terms small), whereas SDW1 polarises strongly for $H \parallel {\bf a}$ (significant $\delta_{1a}$, the relative intensity of the second harmonic against field is shown in Fig.~\ref{Fig_2}(j)). The lack of a 3$^\text{rd}$-order harmonic for SDW1 indicates a weaker role of CEF anisotropy ($\delta_2$ small) in this state compared with SDW2.  \\
\indent Unequal moments along the $a$ and $b$-axes i.e. $m_a \ne m_b$,  provide another source of anisotropy, that does not result by itself in the generation of higher harmonic reflections. In polarised neutron scattering \cite{Jabbar_2015}, it was found that the ratio of  $m_a$ to $m_b$ was around $3 \pm 0.5$ at lower temperature where SDW2 predominates (for both the 1$^\text{st}$ and 3$^\text{rd}$ harmonics) and $2.5\pm0.5$ (for both SDW1 and SDW2) at higher temperatures. Thus, the intrinsic anisotropy  $m_a$/$m_b$ is similar in all the states. \\
 \begin{figure}[t]
\includegraphics[scale=0.38]{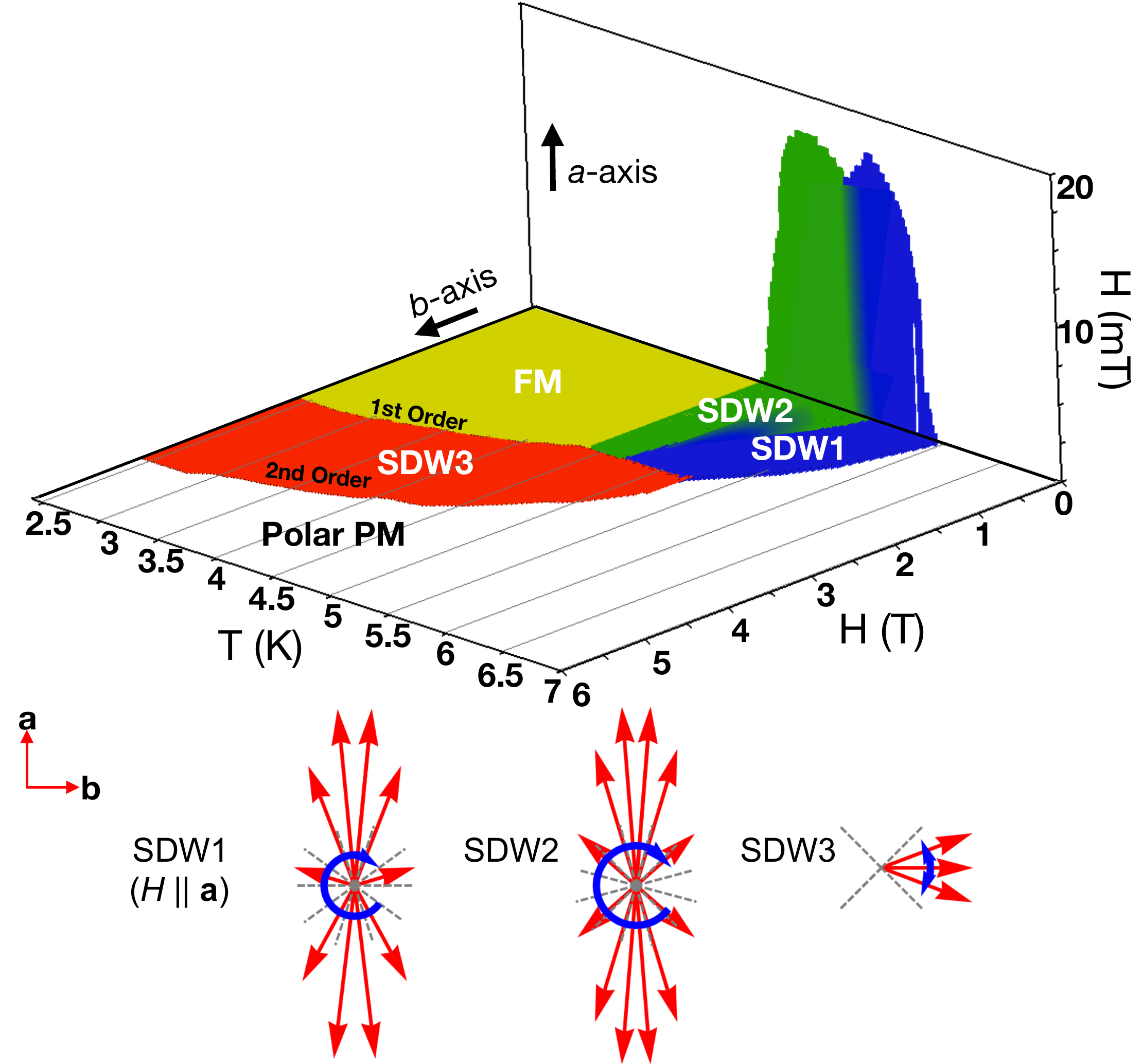}
\caption{The schematic $H_a$-$H_b$-$T$ phase diagram for PrPtAl with fields applied along the easy $a$-axis (vertical) and hard $b$-axis (horizontal), based on our measurements. SDW1 (Blue), SDW2 (green)  and SDW3 (Red) modulated states form a ridge around the first order FM plane (yellow) across which the FM moment ${\bf M} \parallel {\bf a}$ reverses. The phase boundaries between the different modulated states and on the low-field low-temperature side of the ridge are first order.  Schematics of the moment directions viewed along the ${c}$-axis for one modulation period for the different modulated states SDW1, SDW2 and SDW3 are shown below the phase diagram. }
\label{Fig_4}
\end{figure} 
\indent   In forming a modulated state, there is a loss of FM exchange energy since the moments are no longer locally aligned. This acts to minimise the modulation wavevector $q$. This is offset in the QOBD mechanism by the excess density of states created through modulation, roughly proportional to $m^2 q^2$ which lowers the energy, favouring a large $q$. The optimum $q$ results from a subtle balance of these two energies and is strongly dependent on temperature.\\
\indent The crystal field anisotropy energy between the $a$ and $b$ axes is proportional to $m^3$ (or higher power of $m$) and also contributes to the energy balance~\cite{Zener_1954, Jensen}. It favours a state that has all the moments aligned in the preferred CEF direction ($a$-axis). As the temperature increases, the magnitude of the ordered moment falls and the role of magnetic anisotropy decreases more rapidly than the other energy scales. This is consistent with the observed fall in the intensity of the third harmonic with temperature in the SDW2 state.  For  $H\parallel\bf{a}$, the polarisibility should grow with temperature as moments are less confined to the $a$-axis. This is exactly what is seen; for the SDW1 state, where no 3$^\text{rd}$ harmonic is detected,  stronger $H\parallel \bf{a}$ polarisability results in the second harmonic $\delta_{1a}$.   This behaviour is captured in our QOBD model for low applied field, in the helimagnet state.  As in the zero-field case, the  presence of two modulated states, SDW1 and SDW2,  with a jump in $q$ at the transition between them is not found. However,  as shown in SM Fig~S5,  $\delta_1$ becomes dominant over $\delta_2$ on increasing temperature. \\
\indent For larger fields we expect that a fan state around the field direction becomes energetically favorable. We  consider fan states of the form 

\begin{equation}
\mathbf{m}_\textrm{fan}(\mathbf{r}) = \left( \begin{array}{c} m_a\cos\Omega(z) \\   m_b\sin\Omega(z)   \end{array} \right)
\end{equation}

with $\Omega(z) = \Omega_0 +\Delta \sin(qz)$. Here, $\Delta$ is the opening angle of the fan which is centered around the angle $\Omega_0$. We evaluated  the free energy density for $H\parallel\bf{b}$ in the FM, deformed helix and fan states following the QOBD approach. By minimizing the free energies we determined the evolution of the magnetic structure as a function of $H\parallel\bf{b}$, shown in SM Fig.~S3.  Initially, with increasing field,  a deformed helix is favoured that undergoes  a $1^\textrm{st}$ order transition to  a fan state  with a larger value of $q$.  On further increasing the field, this fan continuously transforms to a polarised state.   \\
\indent Our experimental results are, therefore, consistent with the QOBD prediction in applied field. Our resistivity study indeed provides evidence for an increase in the DOS.  Additionally, inelastic neutron scattering shows enhanced magnetic fluctuations, throughout the SDW3 state. The overall $H$-$T$ phase diagram for both hard and easy axes is shown in Fig.~\ref{Fig_4}, along with schematics for each magnetic structure. \\
\indent In conclusion, we have shown PrPtAl may be tuned with field applied along the hard $b$-axis to a singly modulated fan state SDW3, that is well explained by the QOBD theory.   At higher temperature, the low field states SDW1 and SDW2 add complexity. We have resolved why the MR initially increases in SDW1 for $H\parallel\bf{a}$. We have also identified qualitatively why SDW1 is more strongly polarisable and therefore preferred in a small field over SDW2, close to $T_1$. SDW1 and SDW2 are separated by a first order transition and do not necessarily have different symmetry order parameters. Instead they may simply reflect an unstable energy landscape, rather than being intrinsic to the QOBD mechanism. \\
\indent Field induced modulated order has been seen in 1D and 2D materials, including recently in Sr$_3$Ru$_2$O$_7$~\cite{Lester_2015} at the boundary between two field polarised states. While QOBD may contribute to stabilising these modulated states, their low dimensionality means that Van-Hove singularities and nesting probably dominate the formation mechanism.  Non-centrosymmetric  Ca$_3$Ru$_2$O$_7$ provides another interesting example where polarised helicoid order occurs at high temperatures, between two differently oriented anti-ferromagnetic states~\cite{Sokolov_2019,Dashwood_20}. Our results show that modulated state formation may occur more widely when applying a transverse field to a ferromagnet. This could have important implications, for understanding state formation in other materials such as the recently observed field induced superconductivity in UTe$_2$ \cite{Ran_2019_1}. \\
{\bf{Acknowledgements}}; We acknowledge support from the UK EPSRC grants EP/P013686/1 and EP/R013004/1.

\newpage
\pagebreak
\widetext
\setcounter{equation}{0}
\setcounter{figure}{0}
\setcounter{table}{0}
\setcounter{page}{1}
\makeatletter
\renewcommand{\theequation}{S\arabic{equation}}
\renewcommand{\thefigure}{S\arabic{figure}}
\widetext
\widetext
\begin{center}\large\bf{Supplementary Material; Field Induced Modulated State in the Ferromagnet PrPtAl}
\end{center}
\section{Crystal Growth}
High-quality single crystals (RRR $\approx 75$) were grown from stoichiometric masses of starting materials under ultra high vacuum by  Czochralski pulling from a RF-heated melt in a water cooled crucible.  Before the growth Pr (99.9$\%$ Ames) and Pt (99.995$\%$ Alfa Aesar) were outgassed by ultra high vacuum annealing (Pr and Al (99.999$\%$) were also etched to remove surface oxide).\\

\section{Resonant X-Ray Scattering}
Resonant X-ray diffraction was carried out at the XMAS UK-CRG beamline (BM28), ESRF.    The sample (crystal mosaic FWHM $0.01^\circ$) had a natural as grown surface perpendicular to the $c$-axis. Measurements were carried out at the 6.444 keV Pr $L_2$ edge, where a sharp maximum in absorption, fluorescence and magnetic scattering occurs. This comes from a  simple dipole transition, confirmed from an azimuthal scan in zero field (for $q_2=(0,0,2.07)$ at 5K)~[1]. Fields of up to 4~T were applied with a  superconducting magnet (horizontal field transverse to the incident beam), with a  $180^\circ$ vertical and $\pm 5^\circ$ horizontal aperture. The orientation of the sample cryostat can be set independently of the magnet, and  was aligned with the magnet dismounted to give a wider angular coverage, setting the sample $b$-axis parallel to the magnet table with a precision comparable to the sample mosaic.  Measurements were made in the vertical scattering plane, with a LiF (220) analyser to detect scattering in the $\sigma\pi$ channel (Fig.~1, main text). This geometry is sensitive to moments in the $a$$c$ plane.  The error in aligning the $b$-axis with the field is then the larger of $0.01^\circ$  and the precision with which the field is parallel to the magnet axis, estimated to be better than $0.1^\circ$. \\
\indent Measurements with field parallel to the $a$-axis (Fig.~2, main text) were made with the same scattering geometry ($ac$ scattering plane) but the field was applied with a small electromagnet along the $a$-axis and rotated with the sample (the field was determined with a Hall probe). This set-up had a reduced air gap and hence reduced  attenuation from air. \\
\indent Measured satellite intensities have been normalised to the (002) structural Bragg reflection, to allow comparison of intensities between different setups. At the (002) position, the component of the  moment canted along the $c$-axis has a zero structure factor. Thus for both field orientations the measurements made close to this position  are primarily sensitive to the modulated moment along the ${a}$-axis.\\

\section{Inelastic neutron scattering}
\begin{figure}[h]
\includegraphics[scale=0.5]{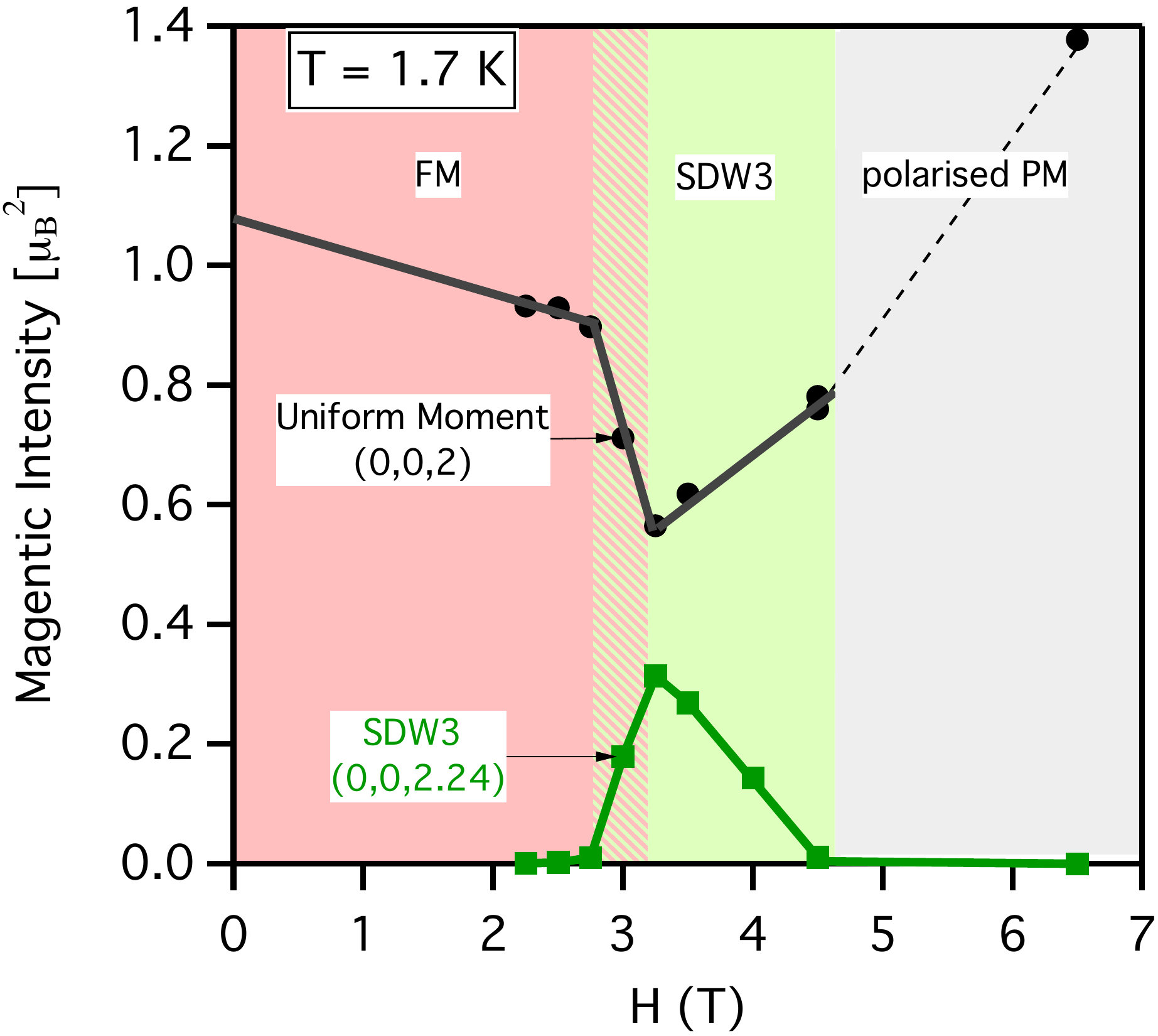}
\caption{The low temperature elastic magnetic intensity  of the (002) and SDW3 $q \approx (0,0,2.24)$  Bragg peaks measured with neutron scattering,   as a function of field $(H)$, applied along the $b$-axis.    }
\label{fig1}
\end{figure}

Inelastic neutron scattering was carried out with the 4F2 cold neutron triple-axis spectrometer at the LLB with   a system of double pyrolytic graphite monochromators and a pyrolytic graphite analyzer. A beryllium filter was put on the final energy arm to remove higher order reflections. The analyzer was kept flat  throughout the experiment, with no horizontal or vertical focusing. No collimator was used.  The experiment was carried out with fixed final neutron energy of $k_f=1.3 \textrm{ \AA}^{-1}$.  The single crystal sample was mounted with [H0L] in the scattering plane  and fields of up to 7 T were applied along the $b$-axis with a superconducting magnet. The sample was mounted on a small goniometer stage and was accurately aligned to have its $b$-axis vertical, parallel with the field.   The diffraction data presented below and the inelastic measurements shown in the main text were measured with the same set-up and experimental conditions. \\
\indent Fig.~\ref{fig1}  shows the low temperature field dependence of elastic magnetic intensity at the (002) Bragg peak. The nuclear contribution to the scattering was measured in zero field just above $T_N$ and subtracted from the total intensity to give the magnetic intensity. The magnetic intensity is proportional to the total uniform moment  squared  $M^2=M_a^2+M_b^2$, expressed in units $\mu_B^2$ by normalising to the nuclear intensity, assuming a free ion Pr$^{3+}$ form factor. Note that the induced moment along the $b$-axis is not fully saturated in fields as high as 25~T~[2].   The scattering intensity of the SDW3 peak at $q \approx (0,0,2.24)$ is also shown in the same intensity units (including a factor of two  to account for multiplicity).  The data clearly shows that the uniform moment is suppressed (due to a reduction of $M_a$) when the SDW3 state is formed in increasing field.  \\

\section{Resistivity}
\begin{figure}[h]
\includegraphics[scale=0.6]{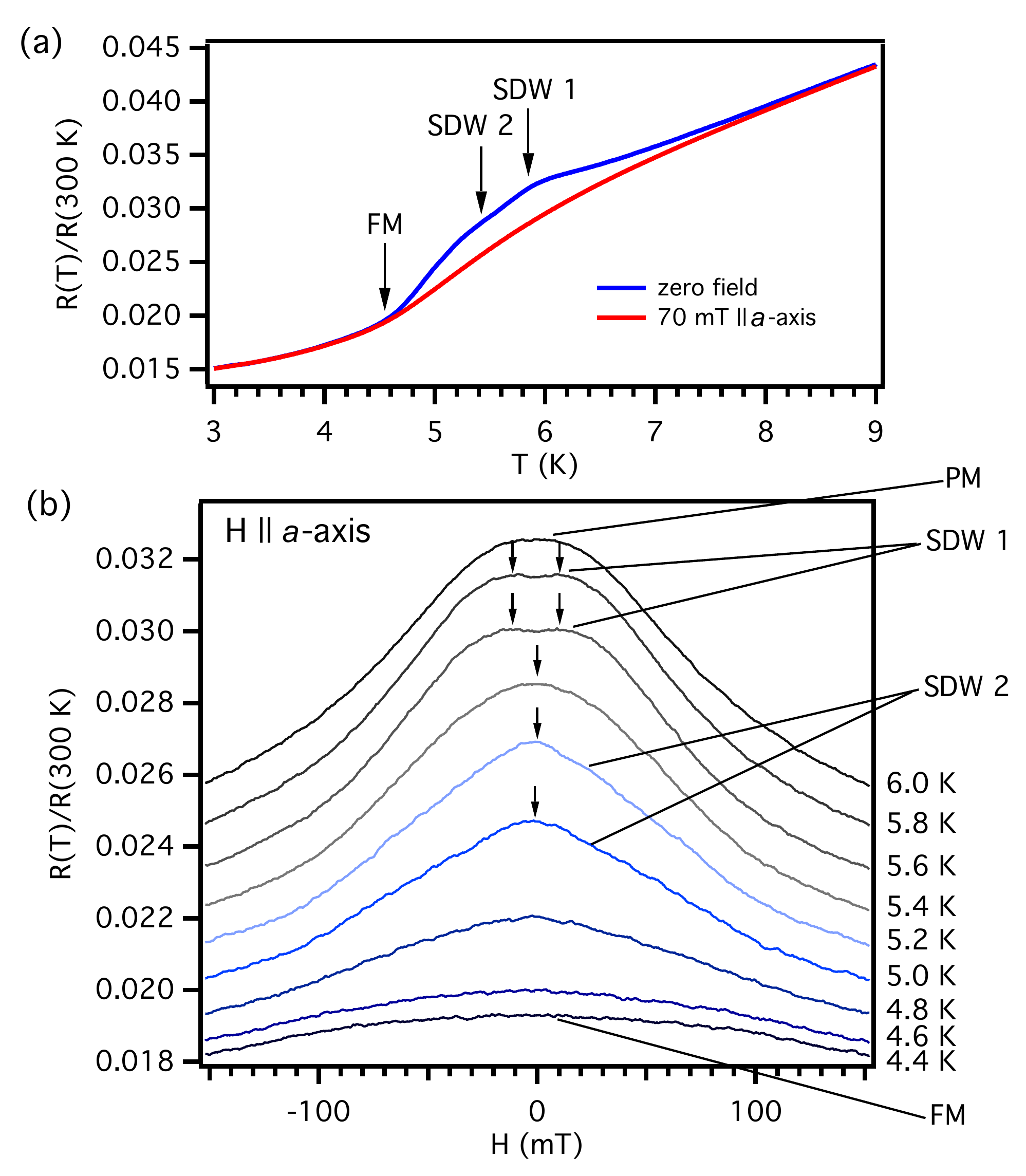}
\caption{(a) Low temperature resistivity of PrPtAl normalised to the value at 300 K,  at zero field (blue) and 70 mT applied along the $a$-axis (red). (b) Magnetoresistance for field applied along the $a$-axis. Arrows indicate the maxima at $\approx 10$ mT in SDW1 and the cusp in SDW2.  Paramagnetic (PM) and ferromagnetic (FM) states are also marked for comparison.       }
\label{fig2}
\end{figure}

 Resistivity and magnetoresistance (MR) measurements  were made with a standard 4 probe technique (100~$\mu$A current  applied along the $c$-axis at 37~Hz).  Experiments were carried out in a $^4$He closed cycle refrigerator with 9~T  superconducting magnet.  The sample was mounted  on a 2-axis rotating stage, fitted with  Hall probes to allow accurate adjustment of the applied field direction, to align  the field precisely along the $b$-axis (Fig.~3, main text).  \\
\indent The resistivity at  zero field,  normalised to the value at 300~K, is shown in Fig.~\ref{fig2}(a). This shows a stronger $T$ dependence in the SDW2 state than below or above it, which provides evidence for an enhanced DOS. The MR, normalised to the zero field resistivity value at 300~K, for field along the easy $a$-axis is shown in Fig.~\ref{fig2}(b).  The low field MR is strongly negative in the SDW2 state with a cusp-like maximum at $H=0$, suggesting stronger fluctuations are present in the SDW2 state than in the FM state, supporting a QOBD based explanation for SDW2. The MR for SDW1 has an initial peak  at low field, that in the current work we show, is a result of  the amplitude of the SDW1 state being initially enhanced with magnetic field.  The structure in resistivity, linked to SDW1 and SDW2, is washed out by field along the  easy $a$-axis and is completely absent for B$>$ 40 mT, consistent with the suppression of these states.\\
 
 \section{Quantum order by disorder in the presence of magnetic anisotropy and field}
 We analyze if the fermionic quantum order-by-disorder mechanism, in the presence of magnetic anisotropy and magnetic field, can qualitatively explain the experimentally observed phase diagram. For simplicity, we consider electronic quasiparticles with an isotropic dispersion $\epsilon(\mathbf{k}) = \hbar^2\mathbf{k}^2/(2m^*)$ and subject to a local repulsion $U$. At the mean-field level, the system undergoes a Stoner transition to a ferromagnetic state at $\rho U_c=1$, where $\rho$ denotes the density of states at the Fermi level. In dimensionless units, the mean-field free energy density is given by $f_\textrm{mf} = \alpha(T)\mathbf{m}^2 + \beta \mathbf{m}^4+\gamma\mathbf{m}^6$, with coefficients $\alpha(T) = 1/(\rho U) -1+\frac{\pi^2}{12} (T/T_F)^2$, $\beta=1/48$ and $\gamma=1/384$.

The coupling to soft electronic particle-hole fluctuations gives rise to a free energy contribution $\delta f_\textrm{fl} =\frac12 (\lambda \rho U)^2 \mathbf{m}^4 \ln [ \mathbf{m}^2 + (T/T_F)^2]$~[3], with a dimensionless constants $\lambda$. This term renders the Stoner transition unstable to fluctuation-induced first-order behavior below a tricritical point $P_c$ at temperature $T_c = T_F\exp[-\beta/(\lambda\rho U)^2]$. In the following, we will only include the fluctuation contribution $\delta \beta_\textrm{fl}(T)= (\lambda \rho U)^2 \ln(T/T_F)$ to the $\mathbf{m}^4$ 
coefficient. This is sufficient if we want to understand the behaviour near $P_c$.

It was pointed out in Ref.~[4], that Landau damping of the order parameter field gives rise to  a negative, non-analytic contribution to the static magnetic susceptibility, rendering the Hertz-Millis-Moriya theory~[5,6] unstable towards first-order behavior and incommensurate order. Such modulated magnetism, is expected, since the resulting deformations of the Fermi surface enhance the phase space for electronic particle-hole fluctuations, a mechanism termed fermionic quantum order-by disorder~[7].  

Starting from the isotropic electronic model, it is possible to self-consistently calculate fluctuations around a helimagnetic state with ordering wavevector $\mathbf{q}$. The resulting gradient terms in the free energy, are proportional to the coefficients of the  $\mathbf{q}=0$ terms~[8,9], 
 
\begin{eqnarray}
f & = &  \frac{1}{V} \int \textrm{d}^3\mathbf{r} \Big\{\alpha  \mathbf{m}^2 + (\beta+\delta\beta_\textrm{fl}) \mathbf{m}^4 +\gamma \mathbf{m}^6   +\frac23  (\beta+\delta\beta_\textrm{fl})(\nabla \mathbf{m})^2 + \gamma \mathbf{m}^2(\nabla \mathbf{m})^2 + \frac35 \gamma (\nabla^2 \mathbf{m})^2\Big\}.
\end{eqnarray}
 The simultaneous sign change of the $\mathbf{m}^4$ and $(\nabla \mathbf{m})^2$ coefficients demonstrates that the fluctuation-driven first order transition is pre-empted by the formation of a helimagnetic state. 

\begin{figure}[t]
 \includegraphics[width=0.5\columnwidth]{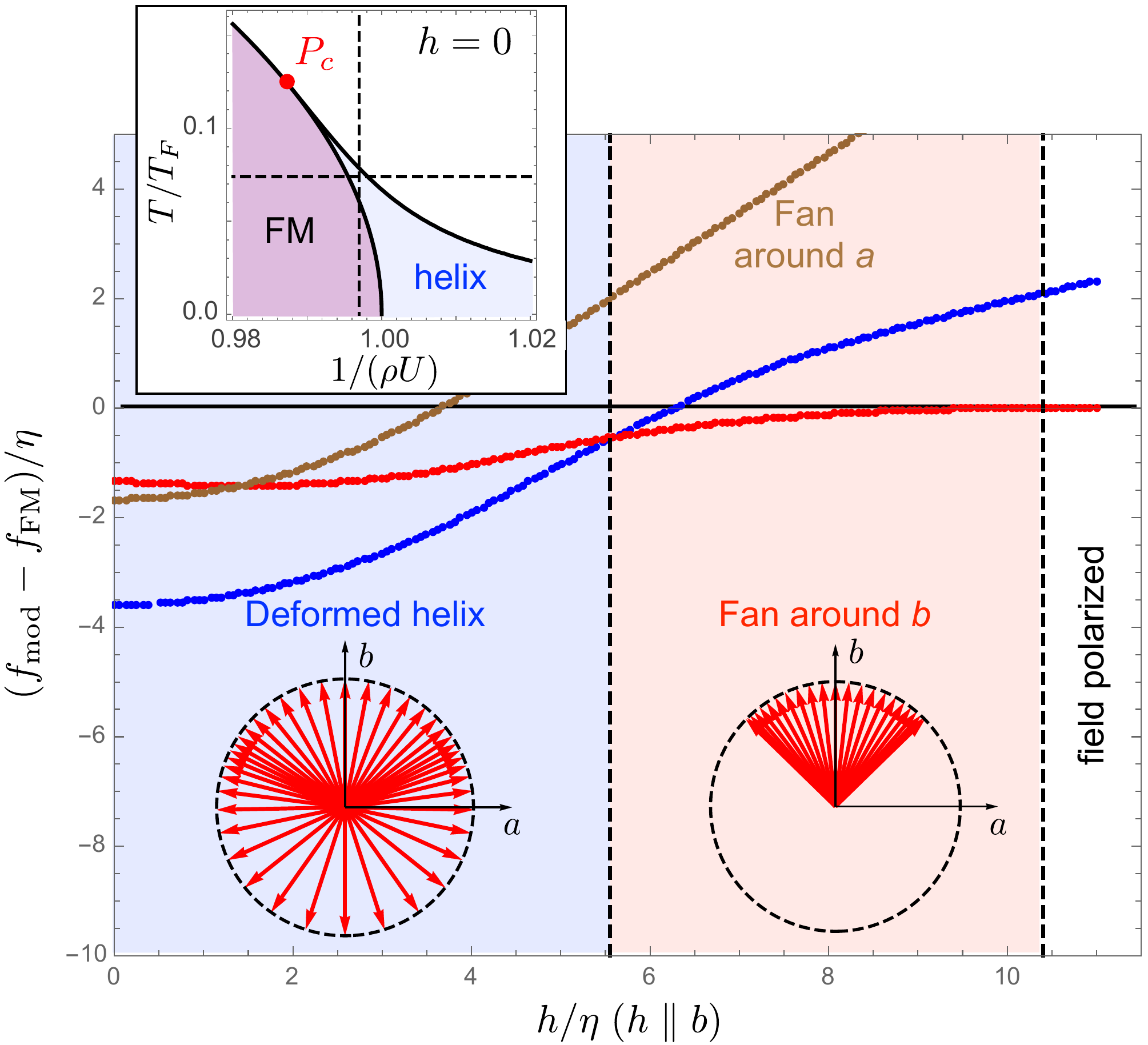}
 \caption{Evolution of the free energies of the fan states and the deformed helix as a function of field along the hard in-plane direction evaluated for $1/(\rho U) = 0.997$ and $T/T_F =0.074$ (these coordinates are shown as a cross hair in the inset phase diagram). The transition from the deformed helix to the 
 field-polarized ferromagnet occurs through an intermediate fan state around the field direction.}
\label{fig3}
\end{figure}

In PrPtAl magnetic anisotropy arises from the coupling of the conduction electrons to local moments. As a result of the $c$-axis being the hard direction, the ordering wavevector of the modulated states is along the $c$-direction and the magnetization confined to the $a$-$b$ plane. In addition, the system exhibits a small in-plane 
anisotropy,
 
\begin{equation}
f_\textrm{ani} = -\frac{\eta}{V} \int \textrm{d}^3\mathbf{r} \;(m_a^2 - m_b^2), 
\end{equation} 
 
with $\eta>0$. Such anisotropy, leads to a deformation of the magnetic helix, resulting in a pronounced third harmonic in the magnetic structure factor. Moreover, since the anisotropy favours the homogeneous ferromagnet over the helimagnet, the region of modulated magnetism is reduced and the transition between the two magnetic phases becomes weakly first order. 
A representative phase diagram for $\lambda=0.1$ and small anisotropy $\eta = 10^{-5}$ (in units of $\rho U^2$) is shown in the inset of Fig.~\ref{fig3} as a function of the inverse dimensionless interaction strength $1/(\rho U)$ and dimensionless temperature $T/T_F$. 

We now investigate the effects of a magnetic field along the hard in-plane direction $\bf{b}$, 
 
\begin{equation}
f_h = -\frac{h}{V} \int \textrm{d}^3\mathbf{r} \;m_b. 
\end{equation} 
 
We expect that the helimagnet is stable for small $h$ and described by the order parameter 
 
\begin{equation}
\mathbf{m}_\textrm{helix}(\mathbf{r}) =  \left( \begin{array}{c} m_a\cos[qz + \phi(z)] \\   m_b\sin[qz + \phi(z) +\epsilon]   \end{array} \right)
\end{equation}
  
with $\phi(z) =-\delta_{1a}\sin(qz) + \delta_{1b} \cos(qz) - \delta_2\sin(2qz)$. Such a deformed helix with $\delta_1,\delta_2>0$ is shown in Fig.~\ref{fig3}. The distortion $\delta_1$ tilts the moments towards the field direction (positive $b$-axis), while $\delta_2$ lowers the anisotropy energy by bunching the moments towards the $\pm a$-axis. Numerical minimization of the free energy shows that higher harmonics of $\phi(z)$ are negligible. We neglect modulations of the amplitude $|\mathbf{m}(\mathbf{r})|$ which are expected to be small. 

For larger fields, one might expect that a fan state around the field direction becomes energetically favourable. We therefore consider fan states of the form 
 
\begin{equation}
\mathbf{m}_\textrm{fan}(\mathbf{r}) =  \left( \begin{array}{c} m_a\cos\Omega(z) \\   m_b\sin\Omega(z)   \end{array} \right)
\end{equation}
  
with $\Omega(z) = \Omega_0 +\Delta \sin(qz)$. Here, $\Delta$ is the opening angle of the fan which is centered around the angle $\Omega_0$. Fan states along the field direction are described by  $\Omega_0=\pi/2$, fan states around the easy $a$ axis by $\Omega_0\in\{0,\pi\}$.

Evaluating the free energy density $f+f_\textrm{ani}+f_h$ for the homogeneous ferromagnet, deformed helix and fan states we obtain

\begin{align}
&  f_\textrm{FM} (m_a,m_b)  =  \alpha (m_a^2+m_b^2) + \tilde{\beta} (m_a^2+m_b^2)^2  +\gamma (m_a^2+m_b^2)^3 -\eta (m_a^2-m_b^2)-h \,m_b \\
& f_\textrm{helix} (m,q,\delta_1,\delta_2)  =  \alpha m^2 + \tilde{\beta} m^4+\gamma m^6 +\frac23 \tilde{\beta} q^2m^2  + \frac35 \gamma q^4 m^2 +\gamma q^2 m^4 +\frac{3}{10}\gamma q^4 m^2 (\delta_1^2+4\delta_2^2)\nonumber \\
& \qquad -\eta \,m^2 \left[J_1(2\delta_2)-J_2(2\delta_1) -\delta_1^2 \delta_2\right] -h\,m \left[J_1(\delta_1)+ \frac14 \delta_1\delta_2 -\frac18 \delta_1\delta_2^2  \right]\\
& f_\textrm{fan} (m,q,\Omega_0,\Delta)  = \alpha m^2 + \tilde{\beta} m^4+\gamma m^6 +\frac13 \tilde{\beta}  \Delta^2 q^2m^2 + \frac{3}{10} \gamma\Delta^2\left(1+\frac34 \Delta^2\right)q^4m^2+\frac12 \gamma \Delta^2 q^2 m^4\nonumber\\
& \qquad  -\eta \,m^2\cos(2\Omega_0) J_0(2\Delta)-h\,m \sin(\Omega_0)J_0(\Delta)
\end{align}
 
where $J_0$, $J_1$ and $J_2$ denote Bessel functions of the first kind and $\tilde{\beta} = \beta+\delta\beta_\textrm{fl}$.

We minimize the free energies for the different states and identify phase transitions from free energy crossings. Since the free energy expansions 
are only controlled in the close vicinity of $P_c$, where the magnetization $m$ and the ordering wavevector $q$ are small, it is not possible to map out the full phase  diagram. Instead, we fix  $1/(\rho U) = 0.997$ and investigate the behavior along different cuts through the $h$-$T$ phase diagram close to $P_c$.

In Fig.~\ref{fig3}, the free energies of the different modulated states relative to the ferromagnet, $f_\textrm{mod}-f_\textrm{FM}$, are shown as a function of field, $h/\eta$, for a fixed $T/T_F=0.074$. As shown in the inset, at $h=0$ this corresponds to a point in the helimagnetic phase above the ferromagnetic region. In the regime of small fields, the deformed helix and the fan states are favored over the homogeneous ferromagnet, where the deformed helix has the lowest free energy. At very small fields, the fan state around the easy axis ($\pm a$ direction) is energetically favored over the fan around the field direction ($b$ axis). The free energies of the two fan states cross at around $h/\eta\approx 1.4$, where the energy scales associated with the in-plane anisotropy and the applied magnetic field are comparable. At a larger field value of $h/\eta\approx 5.6$, the free energies of the helix and the
fan around the field direction cross, indicative of a first-order transition between the two modulated states. The fan remains the state with the lowest free energy up to a field $h/\eta\approx 10.5$ at which the system becomes fully polarized along the field direction. The free energy of the fan state smoothly approaches that of the polarized ferromagnet, suggesting that the transition is continuous.

\begin{figure}[t]
 \includegraphics[width=0.5\columnwidth]{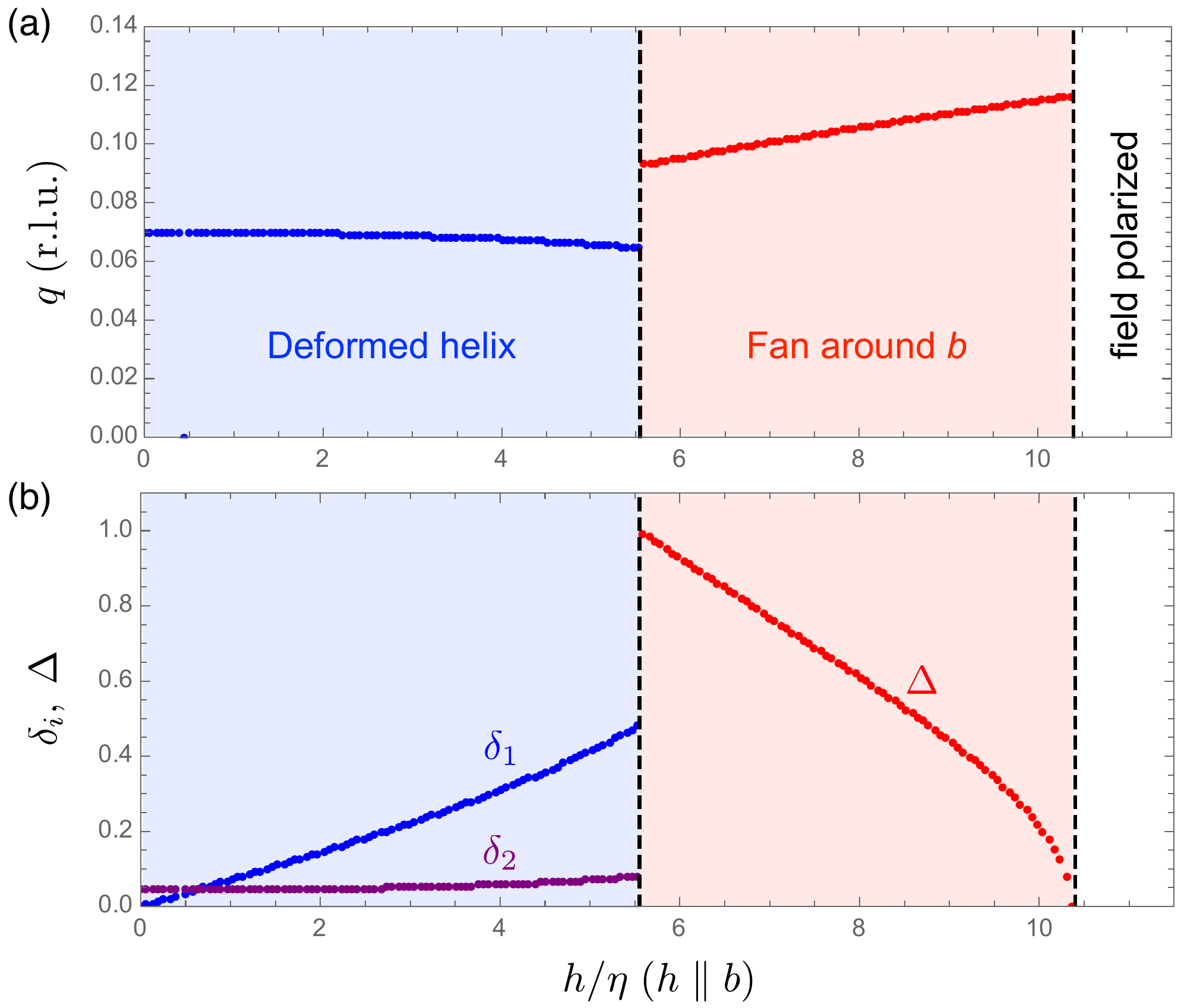}
 \caption{Field dependence of (a) the ordering wavevector $q$ and (b) the deformations $\delta_1$, $\delta_2$ of the helix and the opening angle $\Delta$ of the fan.}
\label{fig4}
\end{figure}

The nature of the phase transitions is evident from the behavior of the order parameters, shown in Fig.~\ref{fig4}. The ordering wavevector $q$ jumps at the first-order transition between the deformed helix and the fan. $q$ has a larger value in the fan state, as seen in experiment. The opening angle $\Delta$ of the fan state has a strong field dependence and goes to zero at the continuous transition to the polarized ferromagnet. As expected, the deformation $\delta_1$ of the helix increases linearly with field, while the distortion $\delta_2$ towards the easy axis is determined by the anisotropy $\eta$ and is almost independent of the applied field. Experimentally $\delta_1$ is smaller than $\delta_2$ over the full field range. This would result from a higher value of $\eta$ than used in the illustrative calculation. Increasing $\eta$ increases $\delta_2$ and this in turn reduces the rate of increase of $\delta_1$ with field. 

\begin{figure}[t]
 \includegraphics[width=0.5\columnwidth]{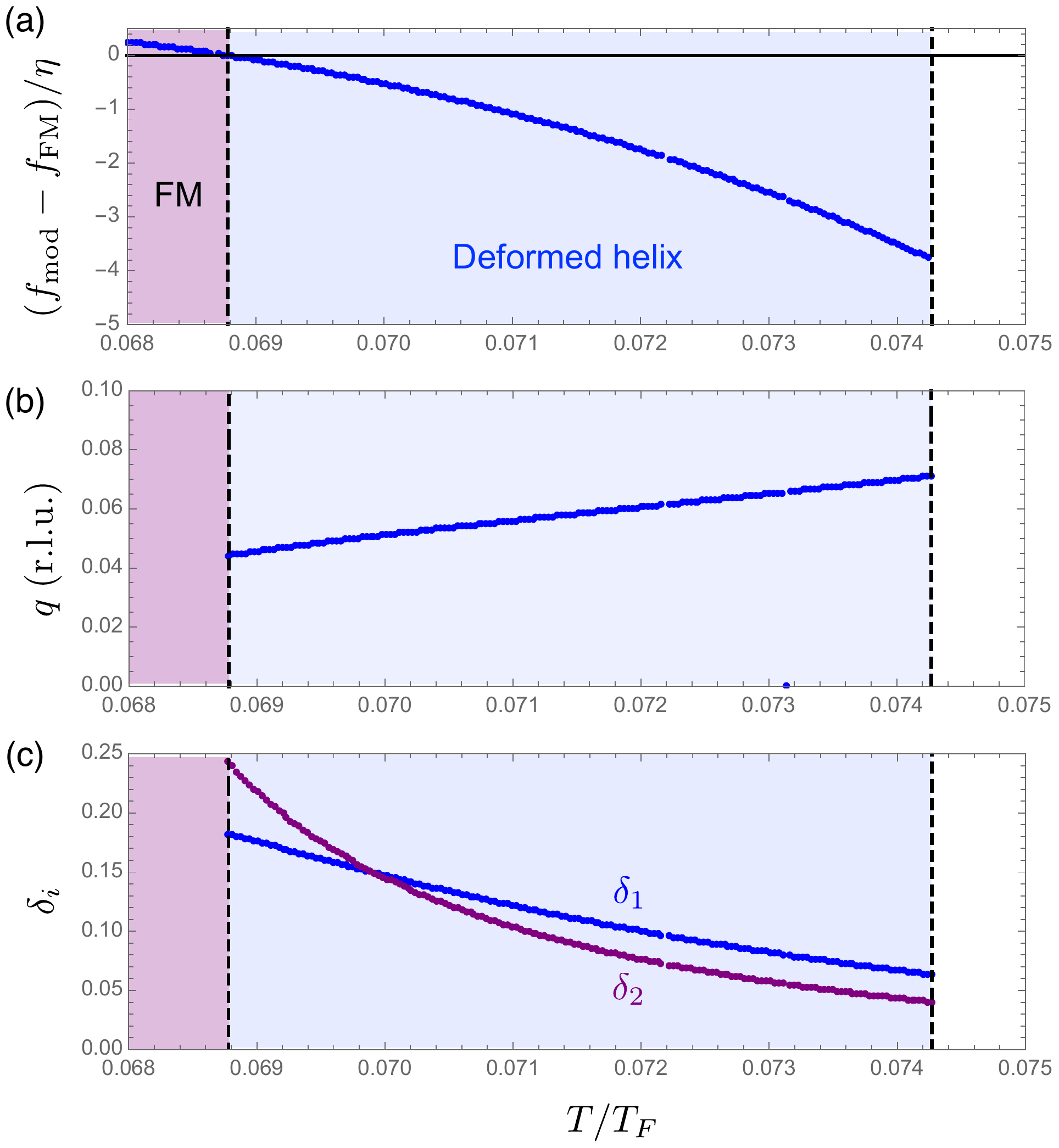}
 \caption{Temperature dependence of (a)  the free energy, (b) the ordering wavevector $q$ and (c) the deformations $\delta_1$ and $\delta_2$ of the deformed helix state at fixed field. }
\label{fig5}
\end{figure}

We now investigate the temperature dependence at fixed field $h/\eta = 0.95$, well below the critical field where the first-order transition to the fan state occurs. At such small fields the ferromagnet is stable at low temperatures. The free energy difference $f_\textrm{helix}-f_\textrm{FM}$ (Fig.~\ref{fig5}(a)) shows a first-order transition from the ferromagnet to the deformed helix at $T/T_F\approx 0.0688$. As shown in Fig.~\ref{fig5}(b), $q$ jumps from zero to a finite value at the transition and increases with temperature.

As in the zero-field case, we don't see any indication for the presence of two modulated states, SDW1 and SDW2, with a jump in $q$ at the transition between the two modulated states.  This could be be due to the simplicity of our model which neglects lattice effects, does not correctly account for the electronic  bandstructure of PrPtAl, and  incorporates local moments only as source of anisotropy. 

 Nevertheless, the temperature dependence of the deformations $\delta_1$ and $\delta_2$ (Fig.~\ref{fig5}(c))  reflects some of the experimentally observed behavior. At low temperatures, $\delta_2$ dominates, giving rise to a strong third harmonic in the magnetic structure factor. As temperature is increased, $\delta_2$ decreases faster than $\delta_1$. Close to the transition to the paramagnet, $\delta_1$ dominates, giving rise to a pronounced second harmonic.   

\clearpage 

\clearpage

\clearpage

\clearpage

\clearpage

\clearpage

\noindent[1] G. Abdul-Jabbar, D. A. Sokolov, C. D. O'Neill, C. Stock, D. Wermeille, F. Demmel, F. Kr\"uger, A. G. Green, F. L\'evy-Bertrand, B. Grenier, et al., Nat. Phys. $\bf{11}$, 321 (2015).

\noindent[2] S. Kato, H. Kitazawa, H. Abe, N. Tsujii, and G. Kido, Physica B: Condensed Matter $\bf{294}$-$\bf{295}$, 217 (2001).

\noindent[3] D. Belitz, T. R. Kirkpatrick, and T. Vojta, Phys. Rev. Lett. $\bf{82}$, 4707 (1999).

\noindent[4] A. V. Chubukov, C. P\'epin, and J. Rech, Phys. Rev. Lett. $\bf{92}$, 147003 (2004).

\noindent[5] J. A. Hertz, Phys. Rev. B $\bf{14}$, 1165 (1976).

\noindent[6] A. J. Millis, Phys. Rev. B $\bf{48}$, 7183 (1993).

\noindent[7] A. G. Green, G. Conduit, and F. Kr\"uger, Annual Review of Condensed Matter Physics $\bf{9}$, 59 (2018).

\noindent[8] U. Karahasanovic, F. Kr\"uger, and A. G. Green, Phys. Rev. B $\bf{85}$, 165111 (2012).

\noindent[9] C. J. Pedder, F. Kr\"uger, and A. G. Green, Phys. Rev. B $\bf{88}$, 165109 (2013).

\end{document}